\documentclass[12pt,preprint]{aastex}

\slugcomment{Draft version from \today, Accepted for ApJ Letter}

\shorttitle{SMA multi-line observations in IRAS\,18089-1732}
\shortauthors{Beuther et al.}

\begin{document}

\title{SMA multi-line observations of the massive star-forming
region IRAS\,18089-1732}


\author{H. Beuther\altaffilmark{1}, 
Q. Zhang\altaffilmark{1}, 
T. Hunter\altaffilmark{1}, 
T.K. Sridharan\altaffilmark{1}, 
J.-H.  Zhao\altaffilmark{1}, 
P. Sollins\altaffilmark{1}, 
P.T.P. Ho\altaffilmark{1,2}, 
S.-Y. Liu\altaffilmark{2}, 
N. Ohashi\altaffilmark{2}, 
Y.N. Su\altaffilmark{2}, 
J. Lim\altaffilmark{2}}

\altaffiltext{1}{Harvard-Smithsonian Center for Astrophysics, 60 Garden Street, Cambridge, MA 02138}
\altaffiltext{2}{Academia Sinica Institute of Astronomy and Astrophysics,  No.1, Roosevelt Rd, Sec. 4, Taipei 106, Taiwan, R.O.C.}

\begin{abstract}
Submillimeter Array (SMA) observations of the high-mass star-forming
region IRAS\,18089-1732 in the 1\,mm and 850\,$\mu$m band with 1\,GHz
bandwidth reveal a wealth of information. We present the observations
of 34 lines from 16 different molecular species. Most molecular line
maps show significant contributions from the outflow, and only few
molecules are confined to the inner core. We present and discuss the
molecular line observations and outline the unique capabilities of the
SMA for future imaging line surveys at high spatial resolution.
\end{abstract}

\keywords{star: formation -- submillimeter -- techniques:
interferometric -- astrochemistry -- line: identification -- ISM:
individual(IRAS\,18089-1732)}

\section{Introduction}

(Sub-)millimeter Interferometry offers great opportunities to study
massive star-forming regions in detail. High-mass star formation
proceeds in a clustered mode, and at the typical distances of a few
kpc high angular resolution is necessary to resolve the different
phenomena taking place. Furthermore, massive star-forming cores emit
strongly in the (sub-)mm regime and exhibit a forest of molecular line
transitions tracing various physical aspects (e.g.,
\citealt{schilke1997b}). Here, we present early results obtained in
the field of massive star formation with the Submillimeter Array
(SMA\footnote{The Submillimeter Array is a joint project between the
Smithsonian Astrophysical Observatory and the Academia Sinica
Institute of Astronomy and Astrophysics, and is funded by the
Smithsonian Institution and the Academia Sinica.}) on Mauna
Kea/Hawaii. Before the advent of the SMA, the only sub-mm
interferometry ever conducted was single-baseline work with the JCMT
and CSO, which aimed largely at low-mass sources because the expected
structures would be simple (e.g., \citealt{brown2000}).

The target source IRAS\,18089-1732 is part of a sample of 69 High-Mass
Protostellar Objects selected mainly via infrared color-color criteria
and the absence of strong cm continuum emission \citep{sridha}. It is
approximately at a distance of 3.6\,kpc\footnote{The kinematic
distance ambiguity is solved by associating the region via the near-
and mid-infrared surveys 2MASS and MSX on larger scales with sources
of known distance (Bontemps, priv. comm.).} and its bolometric
luminosity is about $10^{4.5}$\,L$_{\odot}$ \citep{sridha}. Previous
Millimeter continuum observations reveal a massive core
$>2000$\,M$_{\odot}$ with H$_2$O and CH$_3$OH maser emission, and a
weak 1\,mJy cm source is detected
\citep{beuther2002a,beuther2002c}. As part of a single-dish CO outflow
study, wing emission indicative of molecular outflows was detected
\citep{beuther2002b}, and \citet{sridha} report the detection of the
hot-core molecules CH$_3$CN and CH$_3$OH. Comprising all features of a
young luminous massive star-forming region at an early evolutionary
stage just beginning to form a hot core, IRAS\,18089-1732 is an ideal
target for early science with the SMA. While this letter focuses on
the multi-line observations of IRAS\,18089-1732, an accompanying
letter discusses the (sub-)mm continuum and SiO/HCOOCH$_3$ line data
revealing the disk-jet system (Beuther et al., this volume).

\section{The Submillimeter Array (SMA)}
\label{obs}

IRAS\,18089-1732 was observed with the SMA between May and July 2003
in two different configurations with 3 to 5 antennas in the array\footnote{For more details on the SMA and its specifications see Ho,
Moran \& Lo (this volume) and the SMA web-site:
http://sma-www.harvard.edu/.}. 
The phase reference center of the
observations was R.A.[J2000] 18:11:51.4 and Dec.[J2000]
$-17$:31:28.5. The projected baselines ranged from 10.8/18.6 to 120\,m
for the 1.3\,mm/850\,$\mu$m observations, respectively. For bandpass
calibration we used the planet Mars. The flux density scale was
derived by observations of the planet Uranus and is estimated to be
accurate within 25\,$\%$. Phase and amplitude calibration was done via
frequent observations of the quasar NRAO530, approximately
$10^{\circ}$ from the target source ($S_{\rm{217GHz}}\sim 2.4$\,Jy and
$S_{\rm{354GHz}}\sim 1.5$\,Jy). The zenith opacity for the 350\,GHz
observations~-- measured at 225\,GHz with the NRAO tipping radiometer
operated by the CSO ($\tau(354\,\rm{GHz})\sim
3.2\,\tau(225\,\rm{GHz})$)~-- was excellent for one track
($\tau$(354\,GHz)$\sim 0.19$) but somewhat worse for the second track
($\tau$(354\,GHz)$\sim 0.32$). The data of the second track could only
be used for the 850\,$\mu$m continuum because the signal-to-noise
ratio in the line data was too low. For the 217\,GHz observations the
opacity was slightly higher with $\tau$(225\,GHz)$\sim 0.12$ in one
night and $\tau$(225\,GHz)$\sim 0.3$ in the other night. The
doppler-tracking center frequency on the source was
$v_{\rm{lsr}}=33.8$\,km\,s$^{-1}$. The receivers operate in a
double-sideband mode (DSB) with an IF band of 4--6\,GHz.  The
correlator bandwidth at the time of observation was 1\,GHz, and the
frequency resolution was 0.825\,MHz. In the 850\,$\mu$m band we tuned
the receivers to the HCN(4--3) line at 354.5055\,GHz in the upper
sideband. At 1\,mm the tuning was centered on the SiO(5--4) line at
217.1049\,GHz in the lower sideband. Because of double-sideband
reception we obtained simultaneously data at 344\,GHz and
227\,GHz. The beam sizes of the combined datasets at 1.3\,mm was
$2.7''\times 1.7''$ and at 850\,$\mu$m $1.4'' \times 0.9''$ (uniform
weighting), the beam size of the 850\,$\mu$m line data using only one
track was $1.6'' \times 0.7''$ . System temperatures (DSB) in the
850\,$\mu$m band were between 300-900\,K and in the 1\,mm band around
200\,K. The primary beam at 217\,GHz was $\sim 58''$ and at 354\,GHz
$\sim 36''$. The continuum rms was $\sim 7$\,mJy at 217\,GHz and $\sim
40$\,mJy at 345\,GHz. The calibration was done with the IDL superset
MIR developed originally for OVRO and adapted for the SMA. The imaging
was performed in MIRIAD.

\section{Results}
\label{imaging}

\begin{figure}[htb]
\includegraphics[angle=-90,width=7cm]{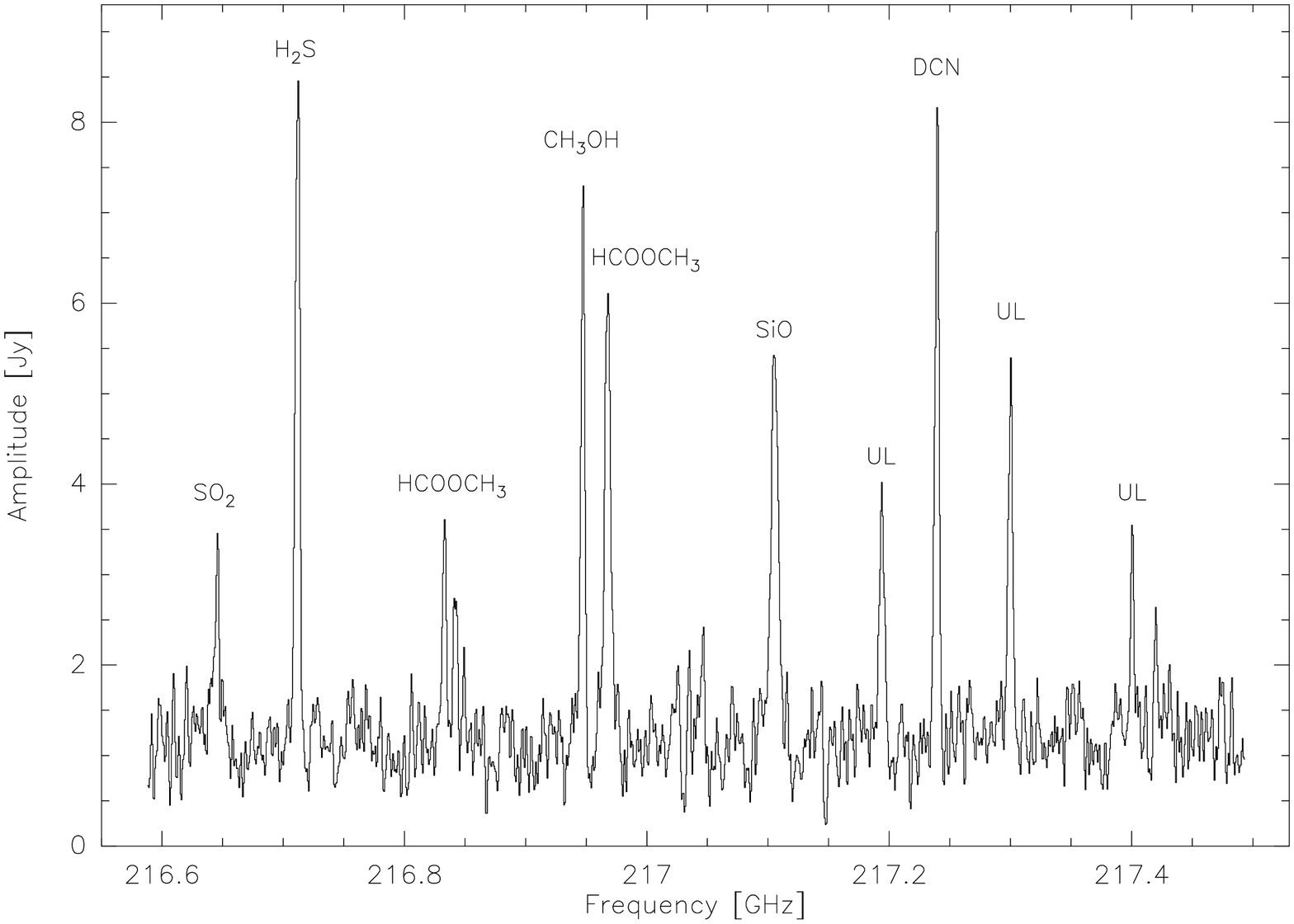}\\
\includegraphics[angle=-90,width=7cm]{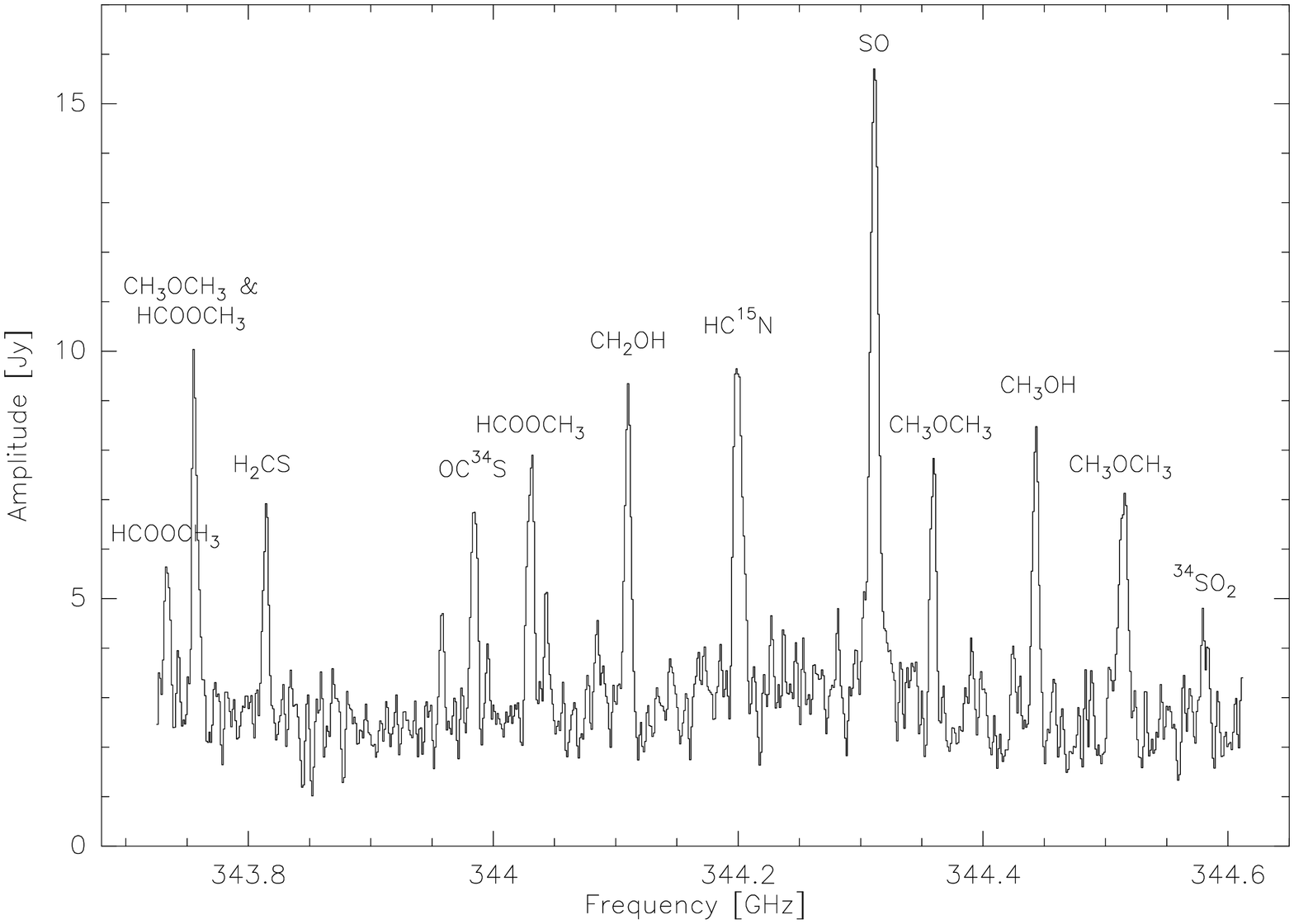}
\caption{SMA spectra toward the high-mass star-forming region
IRAS\,18089-1732. The spectrum on top is taken at 217\,GHz and the bottom
one at 344\,GHz, both with baseline length of approximately 25\,m and
a bandwidth of 1\,GHz. The other sidebands at 227\,GHz and 354\,GHz
are not shown. No continuum subtraction has been performed to show the
different continuum levels in the two bands consistent with dust
emission (see the accompanying paper by Beuther et al., this
volume). UL marks unidentified lines.}
\label{spectra}
\end{figure}

\underline{\it Detection of many molecular lines:} Figure
\ref{spectra} presents two spectra taken at 217\,GHz and 344\,GHz
(both on baselines of $\sim 25$\,m), showing an impressive line forest
of various molecular lines. In the whole dataset with a total
bandwidth of 4\,GHz (1\,GHz in a DSB mode per frequency setup) we
detect 34 molecular lines from 16 molecules/isotopomers, with 5 lines
remaining unidentified (Table \ref{tbl-1}). We detected lines from
molecular species tracing various physical and chemical processes
within massive star-forming regions. While silicon and sulphur bearing
molecules are known to be strong in shocked regions and thus often
trace molecular outflows, we also detected typical hot core molecules
like CH$_3$OCH$_3$, CH$_3$OH, and numerous HCOOCH$_3$ lines. The data
are rich in nitrogen bearing species as HCN, HC$_3$N, and HC$^{15}$N,
which are subject to a different chemistry than, e.g., oxygen bearing
species (e.g., \citealt{wyrowski1999}). Furthermore, we observed the
deuterated species DCN which again is subject to different chemical
networks (e.g., \citealt{hatchell1998a}). It should be noted that the
setup was chosen originally for SiO(5--4) and HCN(4--3), the other
line detections were achieved because of the broad
bandwidth. Transitions of other typical molecules like CH$_3$CN or CN
were absent in this specific setup. The data suggest that because of
the richness of lines in the (sub-)millimeter window, the SMA will
almost always be able to sample many transitions
simultaneously. Figure 1 also shows the different continuum levels at
1\,mm and 850\,$\mu$m of which a detailed analysis is presented in the
accompanying letter (Beuther et al., this volume). We estimate the
line contamination to the broadband continuum flux in the 1.3\,mm band
and the 850\,$\mu$m band both to be about 10\%.

Comparing the line forest with Orion-KL, the proto-typical high-mass
star-forming region toward which many line surveys have been conducted
(e.g., \citealt{sutton1985,schilke1997b}), we detect in the same
frequency ranges all lines previously observed toward Orion-KL, but
with on average significantly narrower line width toward
IRAS\,18089-1732: e.g., toward Orion-KL in the 344\,GHz band the
HC$^{15}$N and CH$_3$OCH$_3$ blend into the broad line wings of the SO
line \citep{schilke1997b} whereas they are clearly resolved from each
other in the IRAS\,18089-1732 data (Fig. \ref{spectra}). The two main
reasons likely contributing to this difference are the larger
luminosity of Orion (10$^5$\,L$_{\odot}$ compared to
10$^{4.5}$\,L$_{\odot}$) and the earlier evolutionary stage of
IRAS\,18089-1732 which is just at the verge of forming a hypercompact
H{\sc ii} region \citep{sridha}.

The large number of observed HCOOCH$_3$ lines allows us to estimate a
temperature for the central region \citep{oesterling1999}. As the data
quality is not good enough to image all lines properly, especially in
the 850\,$\mu$m band (see \S \ref{imaging}), we fitted the intensities
in the visibility domain. From a boltzmann plot we derive a rotation
temperature of $350\pm 100$\,K. As HCOOCH$_3$ traces the dense gas at
the core center, this rather high temperature indicates the existence
of a hot molecular core in the near vicinity of the central
hypercompact H{\sc ii} region \citep{beuther2002c}. Such high
temperatures at the core center are further supported by recent VLA
detections of the NH$_3$(6,6) line with an energy level of 412\,K.

\underline{\it Simultaneous imaging:}
Many of the presented molecular lines are strong enough to image,
which is an enormous step forward compared to single-dish line
surveys. Single-dish observations rarely have sufficient angular
resolution to resolve the sources, although massive star-forming
regions are usually so complex that the chemistry, density and
temperatures exhibit detailed spatial structures. Simultaneous
interferometric imaging of various molecular lines is thus the best
tool to investigate these characteristics in more detail (e.g.,
\citealt{wilner1994,wright1996,blake1996,wyrowski1999}).

Because we have only one track at 850\,$\mu$m with good enough
signal-to-noise ratio to use for the line imaging, we are restricted
in this band to the two strongest lines: HCN and SO (Figure
\ref{images}). In the 1\,mm band we could image many more
lines. Figure \ref{images} presents a characteristic subset of the
integrated emission maps, consisting of the outflow tracer SiO, the
hot-core tracer HCOOCH$_3$, the molecule CH$_3$OH, the sulphur bearing
molecule H$_2$S, the deuterated molecule DCN and the nitrogen bearing
molecule HC$_3$N. Table \ref{tbl-1} presents the peak and integrated
fluxes $S_{\rm{peak}}$ \& $S$. Individual characteristics of the
different species are discussed below. In the accompanying paper
(Beuther et al., this volume), we discuss the SiO(5--4) and
HCOOCH$_3$(20--19) line observations of this dataset in detail. Here
we present only the integrated SiO and HCOOCH$_3$ emission to outline
their outflow/core properties and to associate the emission of the
other molecules with the different physical components within the
region.

\section{Discussion}
\label{discussion}

\begin{figure}[htb]
\includegraphics[width=8.5cm,angle=-90]{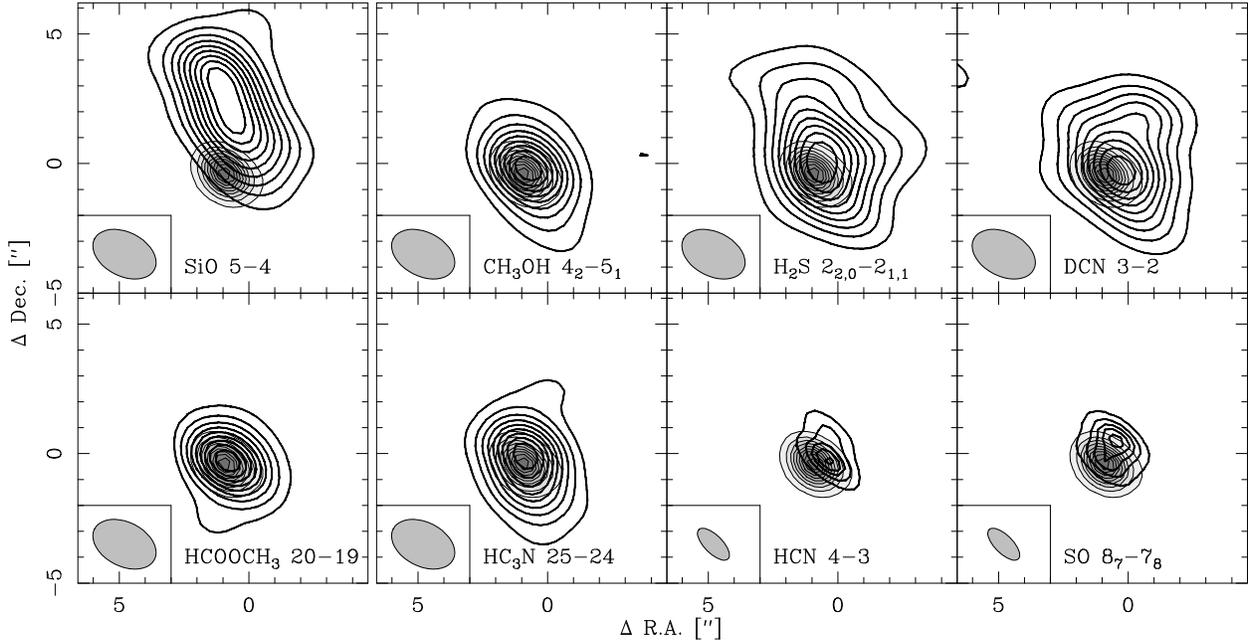}
\caption{Integrated line intensity images of various molecular
transitions toward IRAS\,18089-1732 are shown as thick contours. The
grey-scale and thin contours present the 850\,$\mu$m continuum image
(see also Beuther et al., this volume). The molecules are labeled, and
the beams are shown at the bottom left corner of each panel. The
continuum and the 1\,mm line images are contoured in 10\% level steps
from the integrated intensities presented in Table \ref{tbl-1}. Line
images of HCN and SO are contoured from 10 to 90\% in 20\% steps.}
\label{images}
\end{figure}

As shown in Fig. \ref{images} and outlined in Beuther et al. (this
volume), the SiO(5--4) data depict an outflow emanating from the
central dust core in the northern direction. This outflow is
approximately in the plane of the sky and the SiO emission traces only
the northern lobe. The hot core molecule HCOOCH$_3$(20--19) traces the
central dust condensation with a velocity gradient across the core at
a position angle of $35^{\circ}$ with respect to the outflow
direction. That the HCOOCH$_3$ velocity gradient is not exactly
perpendicular to the outflow can be interpreted as a rotating gas and
dust disk which is influenced by the outflow as well as possible
infall.

\begin{figure}[htb]
\includegraphics[angle=-90,width=7cm]{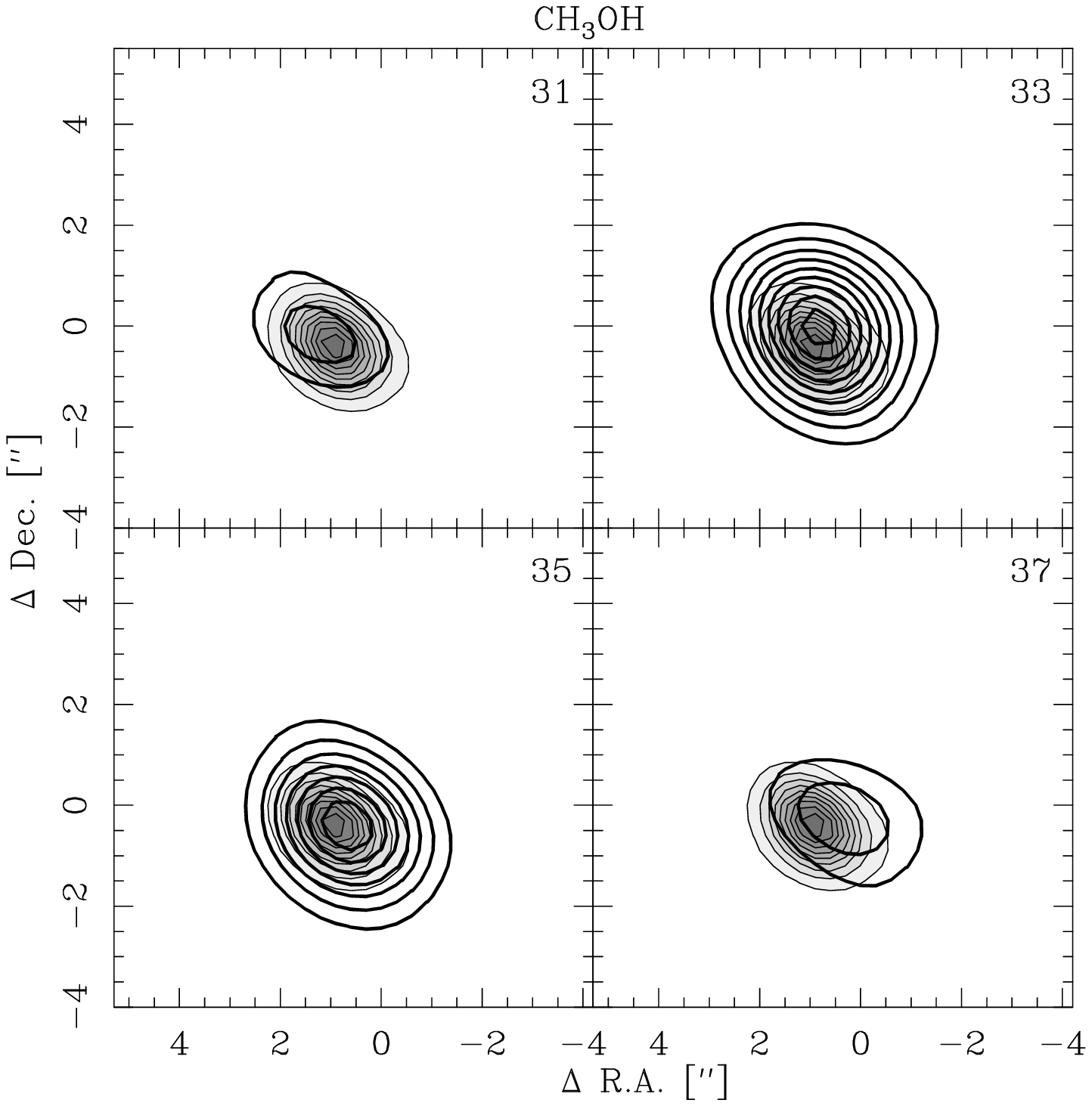}\\
\includegraphics[angle=-90,width=7cm]{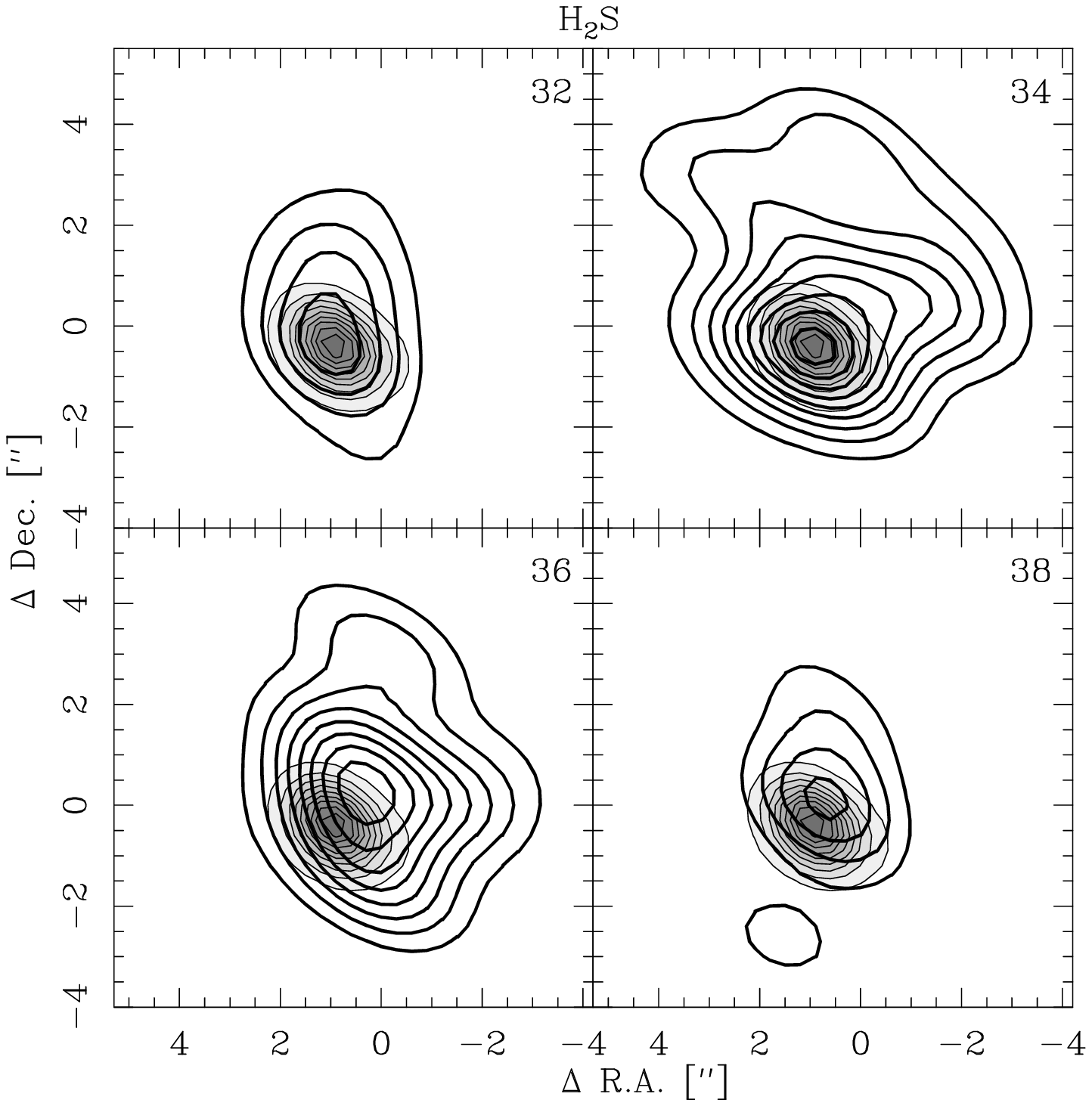}
\caption{Channel maps of CH$_3$OH and H$_2$S. The grey-scale and thin
contours present the 850\,$\mu$m continuum image, the thick contours
show the line emission. The continuum is contoured in 10\% steps from
the peak intensity, and the channel maps are contoured from 15 to 95\%
(10\% steps) from the integrated intensity in the strongest channel,
respectively. The $v_{\rm{lsr}}$ velocity of each channel is marked at
the top-right of each panel.}
\label{channel}
\end{figure}

To give a flavor of the velocity structures varying with molecular
line, Figure \ref{channel} shows two examples of channel maps, with
CH$_3$OH tracing the core and H$_2$S tracing the outflow. The
line-width of the two molecules is not very different because the
outflow is approximately in the plane of the sky. However, the spatial
distribution varies significantly. The core-tracing CH$_3$OH shows a
similar east-west velocity shift as HCOOCH$_3$ whereas the H$_2$S
emission extends to the north comparable to SiO. As the line-widths of
the various molecules are of the same order we base the following
outflow/core associations mainly on morphological
similarities. Comparing the molecular line data shown in Figure
\ref{images} with the results from the SiO/HCOOCH$_3$ observations, we
find that many lines are influenced by the outflow. Most prominently,
this influence is observed in the sulphur bearing species H$_2$S and
SO. However, we find extensions to the north also in DCN and HCN, and,
although to a lesser degree, in HC$_3$N. CH$_3$OH appears to be least
affected by the outflow. Quantitatively speaking, we compare the
source sizes within the $10\%$ levels of the peak emission and the
ratio of integrated to peak flux presented in Table
\ref{tbl-1}. Taking into account the higher angular resolution and the
larger interferometric spatial filtering in the 850\,$\mu$m band,
HCOOCH$_3$, CH$_3$OH and HC$_3$N are rather compact whereas the other
molecules are more extended. This is reflected in the lower integrated
to peak flux ratios of the core-tracing molecules.

These findings are less of a surprise for H$_2$S and SO, which are
known to be released from dust grains in shock interactions with
molecular jets and outflows (e.g., \citealt{bachiller2001}). In
contrast, species like HCN and HC$_3$N were believed to be mainly hot
core tracing molecules \citep{wyrowski1999}. Especially, HCN was
considered as a possible disk tracer in massive star formation, which
is not the case in IRAS\,18089-1732 because the outflow significantly
shapes the overall HCN emission. Similarly, recent observations toward
IRAS\,18566+0408 and IRAS\,20126+4104 reveal strong HCN emission
toward its outflow (Zhang, priv. comm.). The 354\,GHz HCN(4--3) data
need to be improved for a more detailed investigation of HCN outflow
and core emission in IRAS\,18089-1732. On first sight, one gets the
impression that the DCN emission might be more extended than HCN, but
this is mainly due to the different frequency bands and thus varying
interferometric spatial filtering. Therefore, we re-imaged the DCN
data using a short baseline cutoff as given by the HCN observations
($\geq 22$\,k$\lambda$), and we find that the spatial distribution of
HCN and its deuterated species DCN is quite similar.  The compact
CH$_3$OH emission exhibits an east-west velocity gradient similar to
HCOOCH$_3$(20--19). However, CH$_3$OH has been observed to be
associated with outflows in other sources (e.g.,
\citealt{beuther2002d}). Thus, its emission is not always
unambiguously confined to core emission and has to be treated with
similar caution as HCN.

The data shown in this letter outline the unique capacities of the SMA
in performing imaging spectral line surveys. The presented data were
taken in summer time in a testing mode with the partially completed
SMA. Therefore, using the full capabilities of the SMA in the very
near future (e.g., doubled bandwidth of 2\,GHz, 8 antennas within the
array, 2 receivers simultaneously, see also Ho et al., this volume) we
will get a far better sensitivity and imaging quality for forthcoming
physical and chemical studies of various astrophysical objects.

\acknowledgments{We like to thank the whole SMA staff for making this
instrument possible!  H.B. acknowledges financial support by the
Emmy-Noether-Program of the Deutsche Forschungsgemeinschaft (DFG,
grant BE2578/1).}


\begin{thebibliography}{15}
\expandafter\ifx\csname natexlab\endcsname\relax\def\natexlab#1{#1}\fi

\bibitem[{{Bachiller} {et~al.}(2001){Bachiller}, {P{\' e}rez Guti{\' e}rrez},
  {Kumar}, \& {Tafalla}}]{bachiller2001}
{Bachiller}, R., {P{\' e}rez Guti{\' e}rrez}, M., {Kumar}, M.~S.~N., \&
  {Tafalla}, M. 2001, \aap, 372, 899

\bibitem[{{Beuther} {et~al.}(2002{\natexlab{a}}){Beuther}, {Schilke}, {Gueth},
  {McCaughrean}, {Andersen}, {Sridharan}, \& {Menten}}]{beuther2002d}
{Beuther}, H., {Schilke}, P., {Gueth}, F., {et~al.} 2002{\natexlab{a}}, \aap,
  387, 931

\bibitem[{{Beuther} {et~al.}(2002{\natexlab{b}}){Beuther}, {Schilke}, {Menten},
  {Motte}, {Sridharan}, \& {Wyrowski}}]{beuther2002a}
{Beuther}, H., {Schilke}, P., {Menten}, K.~M., {et~al.} 2002{\natexlab{b}},
  \apj, 566, 945

\bibitem[{{Beuther} {et~al.}(2002{\natexlab{c}}){Beuther}, {Schilke},
  {Sridharan}, {Menten}, {Walmsley}, \& {Wyrowski}}]{beuther2002b}
{Beuther}, H., {Schilke}, P., {Sridharan}, T.~K., {et~al.} 2002{\natexlab{c}},
  \aap, 383, 892

\bibitem[{{Beuther} {et~al.}(2002{\natexlab{d}}){Beuther}, {Walsh}, {Schilke},
  {Sridharan}, {Menten}, \& {Wyrowski}}]{beuther2002c}
{Beuther}, H., {Walsh}, A., {Schilke}, P., {et~al.} 2002{\natexlab{d}}, \aap,
  390, 289

\bibitem[{{Blake} {et~al.}(1996){Blake}, {Mundy}, {Carlstrom}, {Padin},
  {Scott}, {Scoville}, \& {Woody}}]{blake1996}
{Blake}, G.~A., {Mundy}, L.~G., {Carlstrom}, J.~E., {et~al.} 1996, \apjl, 472,
  L49+

\bibitem[{{Brown} {et~al.}(2000){Brown}, {Chandler}, {Carlstrom}, {Hills},
  {Lay}, {Matthews}, {Richer}, \& {Wilson}}]{brown2000}
{Brown}, D.~W., {Chandler}, C.~J., {Carlstrom}, J.~E., {et~al.} 2000, \mnras,
  319, 154

\bibitem[{{Hatchell} {et~al.}(1998){Hatchell}, {Millar}, \&
  {Rodgers}}]{hatchell1998a}
{Hatchell}, J., {Millar}, T.~J., \& {Rodgers}, S.~D. 1998, \aap, 332, 695

\bibitem[{{Oesterling} {et~al.}(1999){Oesterling}, {Albert}, {de Lucia},
  {Sastry}, \& {Herbst}}]{oesterling1999}
{Oesterling}, L.~C., {Albert}, S., {de Lucia}, F.~C., {Sastry}, K.~V.~L.~N., \&
  {Herbst}, E. 1999, \apj, 521, 255

\bibitem[{{Schilke} {et~al.}(1997){Schilke}, {Groesbeck}, {Blake}, \&
  {Phillips}}]{schilke1997b}
{Schilke}, P., {Groesbeck}, T.~D., {Blake}, G.~A., \& {Phillips}, T.~G. 1997,
  \apjs, 108, 301

\bibitem[{{Sridharan} {et~al.}(2002){Sridharan}, {Beuther}, {Schilke},
  {Menten}, \& {Wyrowski}}]{sridha}
{Sridharan}, T.~K., {Beuther}, H., {Schilke}, P., {Menten}, K.~M., \&
  {Wyrowski}, F. 2002, \apj, 566, 931

\bibitem[{{Sutton} {et~al.}(1985){Sutton}, {Blake}, {Masson}, \&
  {Phillips}}]{sutton1985}
{Sutton}, E.~C., {Blake}, G.~A., {Masson}, C.~R., \& {Phillips}, T.~G. 1985,
  \apjs, 58, 341

\bibitem[{{Wilner} {et~al.}(1994){Wilner}, {Wright}, \&
  {Plambeck}}]{wilner1994}
{Wilner}, D.~J., {Wright}, M.~C.~H., \& {Plambeck}, R.~L. 1994, \apj, 422, 642

\bibitem[{{Wright} {et~al.}(1996){Wright}, {Plambeck}, \&
  {Wilner}}]{wright1996}
{Wright}, M.~C.~H., {Plambeck}, R.~L., \& {Wilner}, D.~J. 1996, \apj, 469, 216

\bibitem[{{Wyrowski} {et~al.}(1999){Wyrowski}, {Schilke}, {Walmsley}, \&
  {Menten}}]{wyrowski1999}
{Wyrowski}, F., {Schilke}, P., {Walmsley}, C.~M., \& {Menten}, K.~M. 1999,
  \apjl, 514, L43

\end{thebibliography}

\begin{deluxetable}{lrrrrr}
\tablecaption{Observed lines \label{tbl-1}}
\tablewidth{0pt}
\tablehead{
\colhead{Line} & \colhead{$\nu$} & $S_{\rm{peak}}^b$ & $S^{b,c}$ & Size$^c$ & $\frac{S}{S_{\rm{peak}}}^d$ \\ 
\colhead{} & \colhead{GHz} & $\frac{\rm{Jy}}{\rm{beam}}$ & Jy & $''^2$
}
\startdata
SO$_2$ $22_{2,20}-22_{1,21}$ & 216.643 \\
H$_2$S $2_{2,0}-2_{1,1}$ & 216.710 & 2.3 & 6.4 &39.7 & 0.6 \\ 
HCOOCH$_3$ 18--17 & 216.839 \\			     					 
CH$_3$OH $4_2-5_1$ & 216.946       & 2.0 & 2.8 &20.5 & 0.3 \\ 
HCOOCH$_3$ 20--19 & 216.966         & 2.1 & 2.3 &15.9& 0.2 \\ 
SiO 5--4 & 217.105                  & 1.0 & 2.6 &31.9& 0.6 \\ 
UL$^a$  & 217.193 \\
DCN 3--2 & 217.239                  & 1.9 & 4.3 &32.6& 0.5 \\ 
UL$^a$  & 217.300 \\
UL$^a$  & 217.400 \\
\hline
HCOOCH$_3$ 19--18 E & 227.020 \\		     				 
HCOOCH$_3$ 19--18 A & 227.028 \\		     				 
HC$_3$N 25--24 & 227.419            & 3.4 & 5.9 &21.3& 0.4 \\ 
HCOOCH$_3$ 21--20 & 227.562 \\ 
UL$^a$ & 227.815 \\
\hline	
HCOOCH$_3$ 27--26 E & 343.732 \\
CH$_3$OCH$_3$ 17--16 & 343.753 \\		     				 
HCOOCH$_3$ 27--26 A & 343.758 \\			     					 
H$_2$CS $10_{2,8}-9_{2,7}$ & 343.811\\		     				 
OC$^{34}$S 29--28 & 343.983 \\			     					 
HCOOCH$_3$ 32--31 & 344.030 \\			     					 
CH$_2$OH-E & 344.110 \\				     	
HC$^{15}$N 4--3 & 344.200 \\			     					 
SO $8_8-7_7$ & 344.311             & 2.4 & 4.2 & 5.6 & 1.5 \\ 
CH$_3$OCH$_3$ 19--18 & 344.358 \\		     				 
CH$_3$OH $19_{10,9}-18_{10,8}$ & 344.445 \\	     			 
CH$_3$OCH$_3$ 11--10 & 344.512 \\		     				 
$^{34}$SO$_2$ $19_{1,19}-18_{0,18}$ & 344.581 \\     		 
\hline								 
UL$^a$  & 354.129 \\
$^{13}$CH$_3$OH $4_3-3_2$ & 354.446 \\
HCN(0,1,0) 4--3 & 354.461 \\
HCN 4--3 & 354.506                  & 2.2 & 3.1 & 4.9& 1.3 \\ 
HCOOCH$_3$ 33--32 & 354.608 \\
HC$_3$N 39--38 & 354.699 \\ 
\enddata
\tablenotetext{a}{\footnotesize Unidentified line}
\tablenotetext{b}{\footnotesize Errors of the fluxes are within the calibration uncertainty of $25\%$.}
\tablenotetext{c}{\footnotesize Integrated flux and size within the $10\%$ level of the peak emission}
\tablenotetext{d}{\footnotesize To reduce differences between the two bands (different beams and spatial sampling) this ratio is divided by the beam.}
\end{deluxetable}

\clearpage

\end{document}